\documentclass{osa-article}

\journal{oe}

\begin{document}

\title{Amplification and phase noise transfer of a Kerr microresonator soliton comb for low phase noise THz generation with a high signal-to-noise ratio}

\author{Naoya Kuse,\authormark{1,2,*} and Kaoru Minoshima,\authormark{1,3}}

\address{\authormark{1}Institute of Post-LED Photonics, Tokushima University, 2-1, Minami-Josanjima, Tokushima, Tokushima 770-8506, Japan\\
\authormark{2}PRESTO, Japan Science and Technology Agency, 4-1-8 Honcho, Kawaguchi, Saitama, 332-0012, Japan
\authormark{3}Graduate School of Informatics and Engineering, The University of Electro-Communications, 1-5-1 Chofugaoka, Chofu, Tokyo 182-8585, Japan}

\email{\authormark{*}kuse.naoya@tokushima-u.ac.jp} 



\begin{abstract}
Optical injection locking is implemented to faithfully transfer the phase noise of a dissipative Kerr microresonator soliton comb in addition to the amplification of the Kerr comb. Unlike Er-doped fiber and semiconductor optical amplifiers, the optical injection locking amplifies the comb mode without degrading the optical signal-to-noise ratio. In addition, we show that the residual phase noise of the optical injection locking is sufficiently small to transfer the relative phase noise of comb modes (equivalent to the repetition frequency) of low phase noise Kerr combs, concluding that the optical injection locking of a Kerr comb can be an effective way to generate low phase noise THz waves with a high signal-to-noise ratio through an optical-to-electronic conversion of the Kerr comb.
\end{abstract}

\section{Introduction}
Terahertz (THz) waves with excellent phase noise and large signal-to-noise ratio (SNR), which can support large bandwidth and complex modulation format, are in demand in the era of 5G wireless communication or even beyond (6G) to overcome the exponential growth of data capacity \cite{koenig2013wireless,nagatsuma2016advances,dang2020should}. Electro-optic (EO) frequency combs are one of the effective photonic-based methods to generate THz waves, which overcomes the frequency scaling of electrical oscillators \cite{ducournau2014coherent}. In the method, two sidemodes separated by the integer multiple of the frequency of a microwave oscillator, which drives EO modulators (EOMs), are extracted and converted to THz waves based on optical-to-electronic conversion using a photodetector. A significant drawback of the method is the phase noise multiplication of the microwave oscillator, which indicates the method never overwhelms electrical oscillators in terms of phase noise. On the other hand, optical frequency comb based on mode-locked fiber or solid-state lasers has been used to generate microwaves \cite{fortier2011generation}. Through the optical frequency division (OFD), in which one of the comb modes is phase-locked to a low noise optical reference, microwaves with unprecedented phase noise and frequency stability have been generated \cite{xie2017photonic, nakamura2020coherent}. Even without optical references, extremely low phase noise microwave generation has been demonstrated owing to the intrinsically low phase noise of optical frequency combs \cite{kalubovilage2020ultra}. However, neither fiber nor solid-state combs are suitable for THz waves generation because of the difficulty to multiply low (< 1 GHz) repetition frequency to THz. 

Kerr microresonator soliton comb (hereafter, Kerr comb) \cite{Herr_soliton, kippenberg2018dissipative} with large (> 100 GHz) repetition frequency can generate low phase noise THz waves \cite{zhang2019terahertz, wang2021towards, tetsumoto2021optically}. A value of less than -100 dBc/Hz at a frequency offset of 10 kHz for 1-THz carrier (although measured in the optical domain) \cite{kuse2019control} and 300-GHz carrier \cite{tetsumoto2021optically} as demonstrated would be very difficult to achieve by the method that includes the frequency multiplication.  In addition, as Kerr combs are generated from microresonators fabricated using the CMOS compatible process, systems based on Kerr combs to generate THz waves can be chip-scale and mass productive \cite{xiang2021laser, liu2021high}. Nevertheless, the signal-to-noise ratio (SNR) (equivalent to the white noise) of THz waves generated from Kerr combs is problematic \cite{shafik2006extended,marin2020performance}. The conversion efficiency from a pump laser to Kerr comb is intrinsically small \cite{jang2021conversion}, although several methods to improve the efficiency have been demonstrated \cite{obrzud2017temporal, xue2019super, bao2019laser}, and the optical power of the comb mode ($P_{\rm mode}$) has to be amplified. To generate THz waves, $P_{\rm mode}$ should be more than 10 mW to operate in the saturation regime of a uni-traveling-carrier photodiode (UTC-PD). As $P_{\rm mode}$ is as small as $\sim$ 10 ${\rm \mu}$W before amplification, optical amplifiers can degrade the SNR of comb modes by more than the typical noise figure  (a few dB) of the amplifier, which is transferred to THz waves. Recently, optical injection locking \cite{liu2019optical}, in which one of the comb modes is injected into a single-frequency continuous-wave (CW) laser, has been used to amplify the comb mode of a Kerr comb for wavelength division multiplexing optical communication, showing a gain of 30 dB and optical power of 1 mW \cite{zhang2021low}. However, the most fundamental parameters to generate low noise THz waves, i.e. relative phase noise between the amplified comb modes and residual phase noise between the injected comb mode and single-frequency CW laser, have not been investigated. 

In this study, we show that optical injection locking between a Kerr comb and slave lasers not only amplifies the comb mode, but also faithfully transfers the phase noise of the comb mode to the slave lasers. In the experiments, a comb mode is injected into slave lasers, resulting in an amplification gain of 41 dB (from 2.2 ${\rm \mu}$W to 30 mW) and a locking range of 700 MHz, which is sufficient to infinitely keep optical injection locking without any active feedback loops. Moreover, we show that the residual phase noise of the injection locking when the injection ratio of -44 dB, which is defined as the ratio between the optical power of the injected laser and slave laser is used,  reaches -104, -108, and -114 dBc/Hz at frequency offsets of 10 kHz, 100 kHz, and 1 MHz, respectively. Owing to the small residual phase noise, the phase noise of slave lasers is reduced to the phase noise of the injected comb mode, and the phase noise of the repetition frequency of the Kerr comb is transferred to the relative phase noise between two slave lasers. 

\section{Results}
\subsection{Experimental setup}
\begin{figure}[t!]
\centering\includegraphics[width=13cm]{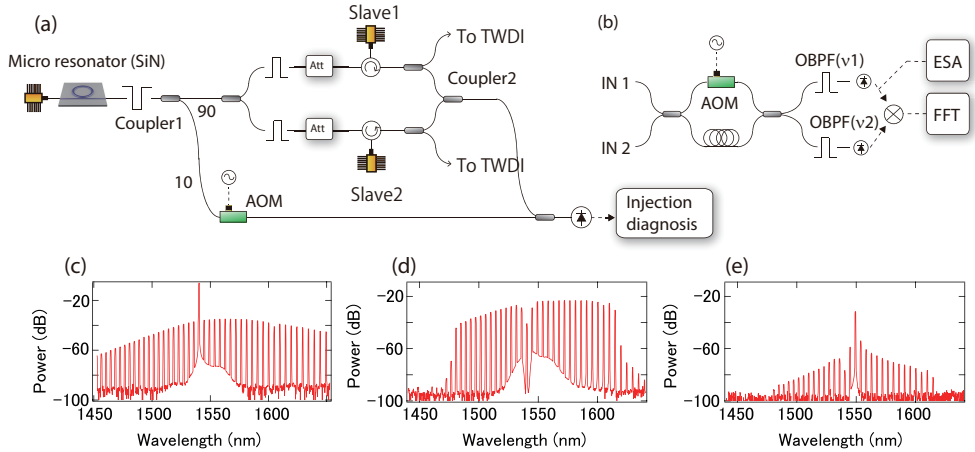}
\caption{(a) Schematic of the experimental setup. Att: attenuator, TWDI: two-wavelength delayed self-heterodyne interferometer. (b) Schematic of the TWDI. OBPF: optical bandpass filter. (c) Optical spectrum of the Kerr comb.  (d) Optical spectrum of the Kerr comb after the notch filter. (e) Optical spectrum of the Kerr comb after the optical bandpass filter. RBW is 1 nm for (c) - (e)}
\end{figure}
Figure 1(a) shows the experimental setup of injection locking. An external-cavity diode laser (ECDL) oscillating at 1540 nm is used as a pump CW laser. The pump CW laser after passing through a dual-parallel Mach-Zehnder modulator (DP-MZM) and an erbium-doped fiber amplifier (EDFA) is coupled into a Si$_3$N$_4$ (SiN) microresonator. The FSR of the microresonator is about 560 GHz. A single soliton comb is generated by rapidly scanning the frequency of the pump CW laser using the DP-MZM \cite{kuse2019control}. The residual pump CW laser is rejected by an optical notch filter. The generated soliton comb is split into two by a 90:10 optical coupler (coupler 1): one for the diagnosis of injection locking and other for injection locking. Before injection locking, one of the comb modes is filtered out using an optical bandpass filter (OBPF), moving to a slave laser through an optical circulator. The slave laser is a distributed feedback (DFB) laser (AA 1401 series from Gooch \& Housego, specified linedwidth of 1 MHz and measured side mode suppression ratio of 50 dB). The injected power ($P_{\rm inj}$) is adjusted by a variable optical attenuator before it goes through the optical circulator. The frequency of a slave laser ($\nu_{\rm slave}$) is pre-adjusted to be close to the frequency of the comb mode, followed by a fine tuning via the control of the current driver of the slave laser while measuring an optical beat between the comb mode and slave laser. The frequency of the optical beat is offset by an acousto-optic modulator (AOM) to measure the phase noise by an electrical spectrum analyzer (ESA). In this study, the Kerr comb is used for two injection lockings to investigate the relative phase noise between the two injection-locked slave lasers ($L_{\rm s-s}$) (slaves 1 and 2 in Fig. 1(a)). The wavelengths of slaves 1 and 2 are 1550 and 1558 nm, respectively, which correspond to the -2nd and -4th comb modes with respect to the pump CW laser. $L_{\rm s-s}$ is measured using a two-wavelength delayed self-heterodyne interferometer (TWDI) (Fig. 1(b)) \cite{kuse2017electro, kuse2018photonic}. The TWDI includes an imbalanced Mach-Zehnder interferometer (i-MZI) with an AOM in one of the arms for the heterodyne measurement. The optical path difference of the i-MZI is 100 ns. The two outputs of the i-MZI pass through OBPFs to extract one of the two slave lasers. The phase noise of the slave laser ($L_{\rm slave}$) is measured by transferring the photodetected signal to an ESA. Moreover, by mixing the two photodetected signals followed by Fourier transforming (FFT), $L_{\rm s-s}$ is obtained. 

The optical spectrum of the Kerr comb is shown in Fig. 1(c). The Kerr comb exhibits comb mode spacing equal to the FSR of the microresonator with a spectrum envelope of sech$^2$, indicating the Kerr comb is a single soliton state. The Kerr comb has a residual pump, which is > 20 dB stronger than other comb modes. The optical spectrum after rejecting the residual pump CW laser is shown in Fig. 1(d). The comb modes exhibit SNRs of 40 dB at $\sim$ 1550 nm limited by the amplified spontaneous emission (ASE) from the EDFA and 60 dB at $\sim$ 1580 nm at 1-nm (125-GHz) resolution bandwidth (RBW). The optical spectrum used for injection locking of slave 1 is shown in Fig. 1(e).

\subsection{Locking range and gain}
The locking range is investigated by measuring the frequency of the optical beat ($f_{\rm beat}$) between the comb mode and slave laser while changing $\nu_{\rm slave}$ with the scan rate of less than 1 Hz (Fig. 2(a)). When $f_{\rm beat}$ is within the locking range, it exhibits a plateau, which indicates $\nu_{\rm slave}$ is compelled to be equal to the frequency of the comb mode ($\nu_{\rm mode}$) owing to the injection locking. Figure 2(b) shows the locking range depending on $P_{\rm inj}$, which is estimated from the measurement such as Fig. 2(a). The locking range is proportional to $\sqrt{P_{\rm inj}}$ in theory \cite{liu2019optical}, assuming $P_{\rm slave}$ is constant (the dashed blue curve in Fig. 2(b)). A slight deviation from the theory is likely to derive from the fluctuation of $\nu_{\rm slave}$ and $\nu_{\rm mode}$. The optical  spectrum of the injection locked slave laser with the optical power is shown in Fig. 2(c). $P_{\rm inj}$ of as small as 2.2 ${\rm \mu}$W is amplified to 30 mW (= $P_{\rm slave}$). In addition to amplification, the unwanted comb mode is well suppressed, and a tiny residual is observed, which is 50 dB smaller than the main carrier. As a result, the SNR of the carrier is improved from 40 (22 dB against the largest comb mode) to 50 dB at 1-nm (125-GHz) RBW with a gain of 41 dB and an output of 30 mW. 

\begin{figure}[t!]
\centering\includegraphics[width=13cm]{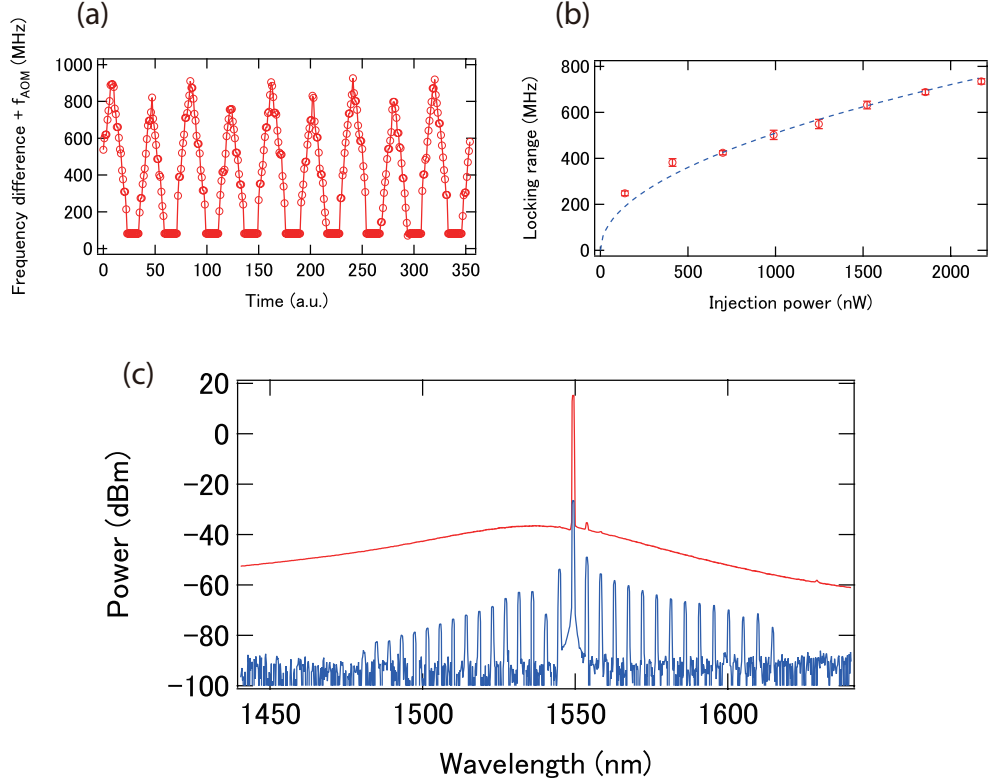}
\caption{(a) Frequency of the optical beat between the slave laser and nearest comb mode when $\nu_{\rm slave}$ is repetitively swept. (b) Range of the injection locking (the red circle) depending on $P_{\rm inj}$. The dashed blue curve shows a fitting curve that is proportional to $\sqrt{P_{\rm inj}}$. (c) Optical spectra of the injection-locked slave laser (red) and the injected comb mode (blue). RBW is 1 nm.}
\end{figure}

\subsection{$L_{\rm slave}$, $L_{\rm residual}$, and $L_{\rm s-s}$}
\begin{figure}[b!]
\centering\includegraphics[width=13cm]{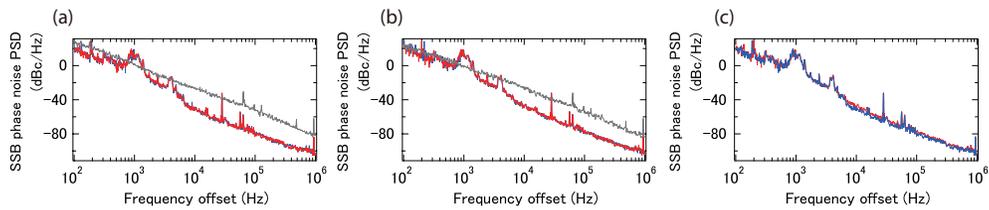}
\caption{(a) Single sideband (SSB) phase noise power spectrum density (PSD) of the injected comb mode at $\sim$ 1550 nm (blue); slave 1 when injection locked (red) and free-running (gray). (b) SSB phase noise PSD of the injected comb mode at $\sim$ 1558 nm (blue); slave 2 when injection locked (red) and free-running (gray). (c) SSB phase noise PSD of the slave 1 (blue) and 2 (red) when injection locked. }
\end{figure}
The phase noise of slave 1(2) ($L_{\rm slave1(2)}$) is compared with that of the injected comb mode ($L_{\rm mode}$). 
As shown in Figs. 3(a) and (b), $L_{\rm slave1(2)}$ is suppressed by $\sim$ 40 dB and is equal to $L_{\rm mode}$ (blue curves in Figs. 3(a) and (b)). $L_{\rm slave2}$ is little higher than $L_{\rm slave1}$ (Fig. 3(c)). This is because $L_{\rm mode}$ is limited by the phase noise of the comb mode spacing probably through the thermorefractive noise of the microresonator, resulting in the larger $L_{\rm mode}$ for slave 2 owing to the larger integer value of the comb mode number with respect to the pump mode \cite{nishimoto2020investigation}. 

Next, the residual phase noise of injection locking ($L_{\rm residual}$) is investigated, in which the phase noise of $f_{\rm beat}$ is measured. $L_{\rm residual}$ is equivalent to an achievable smallest phase noise by the injection locking. 
\begin{figure}[b!]
\centering\includegraphics[width=13cm]{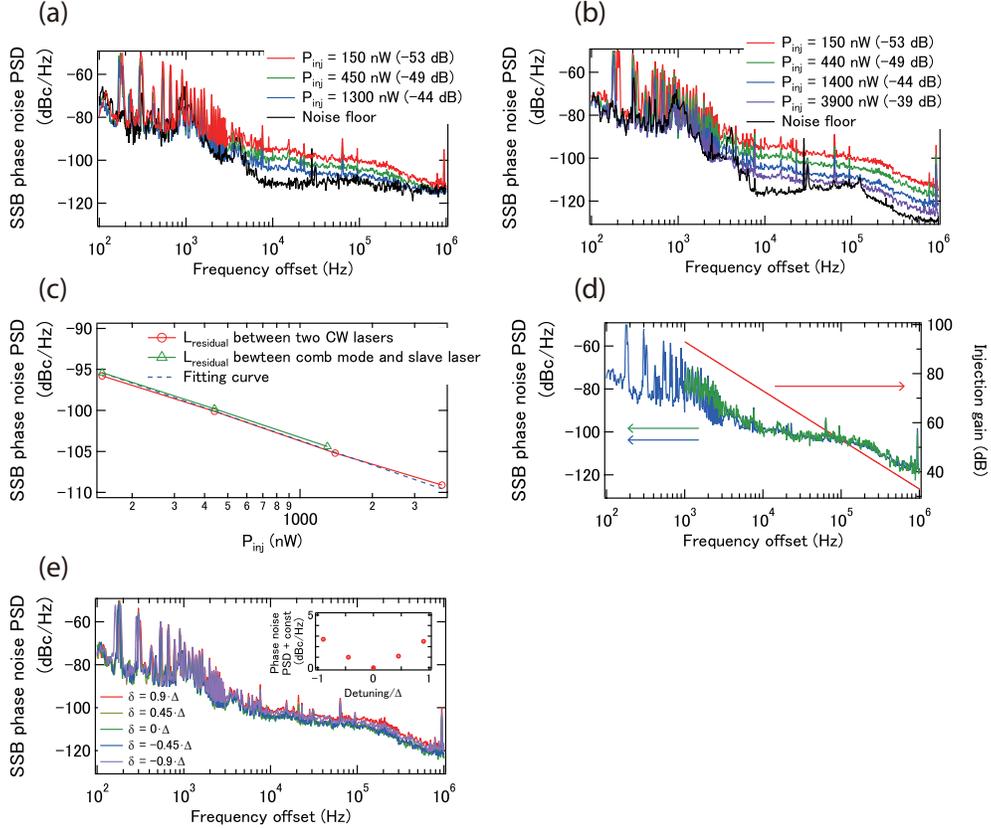}
\caption{(a) SSB phase noise PSD of the optical beat between the injected comb mode and slave laser. (b) SSB phase noise PSD of the optical beat between the injected single-frequency CW laser and slave laser. (c) (Green) SSB phase noise PSD of the optical beat between the injected comb mode and slave laser at a frequency offset of 10 kHz. (Red) SSB phase noise PSD of the optical beat between the injected single-frequency CW laser and slave laser at a frequency offset of 10 kHz with a fitting curve of $-10 \times {\rm log}(P_{\rm inj}) + C$ (the blue dashed line).  (d) (Blue) SSB phase noise PSD of the optical beat between the injected single-frequency CW laser and slave laser when $P_{\rm inj}$ is 150 nW. (Red) An assumed injection gain with $\frac{1}{f^2}$ slope. (Green) SSB phase noise PSD of the free-running slave laser, divided by the injection gain. (e) SSB phase noise PSD of the optical beat between the injected single-frequency CW laser and slave laser when $P_{\rm inj}$ is 150 nW, while varying the detuning between the injected single-frequency CW laser and slave laser. $\delta$ and $\Delta$ denote detuning and the locking range, respectively. Inset shows the SSB phase noise PSD at a frequency offset of 10 kHz . The measurement uncertainty of (a) - (e) is below 1 dB.}
\end{figure}
In the experiments, the difference of the optical path length between couplers 1 and 2 is minimized to reduce the effect of $L_{\rm mode}$ on the measurement, which can set the measurement noise floor. In the first experiment, $L_{\rm residual}$ is measured using the Kerr comb and slave laser (Fig. 4(a)). In this case, the measurement noise floor (the black curve in Fig. 4(a)), which is measured by replacing the slave laser with a fiber mirror, below the frequency offset of 10 kHz is determined by $L_{\rm mode}$. Above the frequency offset of 10 kHz, the white noise of -110 dBc/Hz is set by the SNR of the optical beat. Although the measurement floor is relatively high, the decrease in $L_{\rm residual}$ is observed when $P_{\rm inj}$ increases, which implies that $L_{\rm residual}$ is inversely proportional to $P_{\rm inj}$ (the green curve in Fig. 4(c)). To confirm the scaling in the wide frequency offset, the Kerr comb is switched to a single-frequency CW laser, which provides the lower white noise floor of -130 dBc/Hz (the black curve in Fig. 4(b)). The measurement noise floor within the frequency offset of 10 - 600 kHz is set by the phase noise of the RF oscillator used for the AOM. The same scaling is observed in the wide frequency offset, and $L_{\rm residual}$ reaches -109, -112, and -126 dBc/Hz at frequency offsets of 10 kHz, 100 kHz, and 1 MHz when the injection ratio is -39 dB. In Fig. 4(c), $L_{\rm residual}$ between the injected CW laser and slave laser at a frequency offset of 10 kHz is shown along with a fitting curve of $-10 \times {\rm log}(P_{\rm inj}) + C$ (C: const). According to this scaling, $L_{\rm residual}$ values of -119, -122, and -136 dBc/Hz at frequency offsets of 10 kHz, 100 kHz, and 1 MHz are expected when the injection ratio is -29 dB. The injection ratio of $\sim$ -30 dB is still reasonable for the Kerr comb, because, without extra splitting of the Kerr comb for monitors, $P_{\rm mode}$ of > 10 ${\rm \mu}$ W is feasible. The origin of $L_{\rm residual}$ is the injection gain. In Fig. 4(d), $L_{\rm residual}$ for $P_{\rm inj}$ = 150 nW is shown in addition to the assumed injection gain of $\frac{C}{f^2}$ \cite{ramos1994optical}, where $C$ and $f$ denote constant and frequency, respectively. Dividing $L_{\rm slave}$ for free-running by the assumed injection gain, $L_{\rm residual}$ is overlapped with $L_{\rm slave}$ with the injection gain, which suggests that the use of a low phase noise slave laser can further reduce $L_{\rm residual}$. The dependency of $L_{\rm residual}$ for $P_{\rm inj}$ = 150 nW on the detuning between the injected light and slave laser is measured (Fig. 4(e)). The degradation of $L_{\rm residual}$ over the locking range is as small as 4 dB.

As $L_{\rm residual}$ is far lower than $L_{\rm mode}$, $L_{\rm slave}$ follows $L_{\rm mode}$ (Figs. 3(a) and (b)). However, the situation is different when $L_{\rm s-s}$ is discussed. In Fig. 5, $L_{\rm s-s}$ is compared with the relative phase noise of the comb modes ($L_{\rm relative}$) used for the two injection lockings, which are measured using the TWDI \cite{kuse2017electro, kuse2018photonic} without generating a THz wave through a UTC-PD \cite{wang2021towards}. Although $L_{\rm s-s}$ has well overlapped with $L_{\rm relative}$, a slight deviation from $L_{\rm relative}$ is observed in $L_{\rm s-s}$ at a frequency offset of > 100 kHz. This is because of the insufficient injection gain. Actually $L_{\rm s-s}$ was estimated as $L_{\rm relative}$ + $2^2L_{\rm residual}$ (at $P_{\rm inj}$) (note that this is not expressed in dB, but in linear). Here, $2^2$ accounts for the use of two slave lasers, and the expression is defined in power.

\begin{figure}[h!]
\centering\includegraphics[width=11cm]{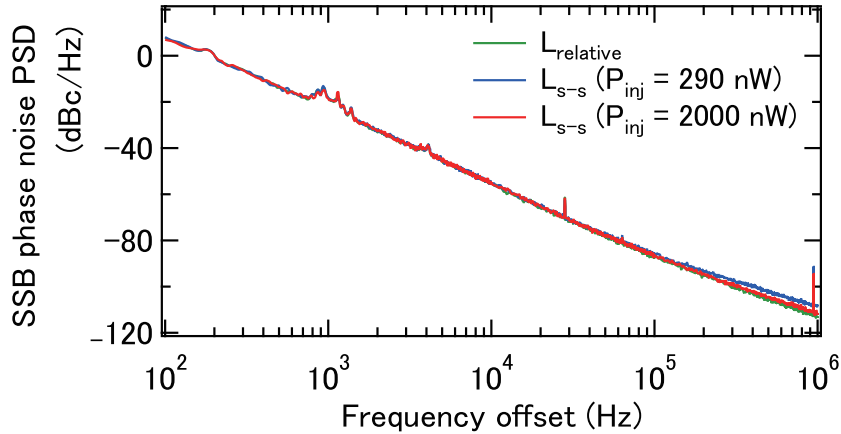}
\caption{(Green) SSB phase noise PSD between the injected comb modes ($L_{\rm relative}$). SSB phase noise PSD between slaves 1 and 2 ($L_{\rm s-s}$) when $P_{\rm inj}$ is 290 (blue) and 2000 nW (red).}
\end{figure}

\section{Conclusion}
We demonstrated optical injection locking between a Kerr comb and slave lasers. Although one comb mode extracted by the OBPF is injected into the slave laser, the OBPF may not be required because the locking range is far smaller than the repetition frequency of the Kerr comb. In addition, the range of the oscillation frequency of the used slave laser is about 200 GHz. Therefore, the slave laser would select one comb mode and be injection-locked to the comb mode without amplifying other comb modes even when all the comb modes would be injected into the slave laser. The obtained SNR of 50 dB (125-GHz RBW), which corresponds to $\sim$ -160 dBc/Hz, is better than the shot noise of UTC-PDs with the saturation electric power of 100 ${\rm \mu}$W. The comb mode is amplified up to 30 mW. When wireless communications with optically generated THz waves are considered, where two optical tones are directed to a UTC-PD, high optical power is preferred owing to the optical loss caused by optical couplers, EOMs, and others. The maximum power of around 100 mW can be achieved by using high-power DFB lasers as slave lasers. In terms of phase noise, we showed that the residual phase noise between the injected comb mode and slave laser is -104, -108, and -114 dBc/Hz at frequency offsets of 10 kHz, 100 kHz, and 1 MHz, respectively, which can be improved by -119, -122, and -136 dBc/Hz at frequency offsets of 10 kHz, 100 kHz, and 1 MHz, respectively, when the injection ratio of $\sim$ -30 dB is used. The residual phase noise is good enough for state-of-the-art low phase noise Kerr combs with a repetition frequency of > 100 GHz. According to these results, we believe Kerr-comb-based THz wireless communications with optical injection locking are very effective in the era of 5G or 6G.

\section*{Funding}
This work was supported by JST PRESTO (JPMJPR1905), Japan Society for the Promotion of Science (21K18726, and 21H01848), Cabinet Office, Government of Japan (Subsidy for Reg. Univ. and Reg. Ind. Creation), Research Foundation for Opto-Science and Technology, KDDI Foundation, Telecommunications Advancement Foundation, and "R\&D of high-speed THz communication based on radio and optical direct conversion" (JPJ000254) made with the Ministry of Internal Affairs and Communications of Japan.
\\
\section*{Disclosures} The authors declare no conflicts of interest.
\section*{Data availability} Data underlying the results presented in this paper are not publicly available at this time but may be obtained from the authors upon reasonable request.

\bibliography{my_reference}






\end{document}